\documentclass[showpacs,a4paper,rmp]{revtex4}
\usepackage{graphicx}
\usepackage{amsmath}
\usepackage{amssymb}

\newcommand{\sus}{susceptibility}

\begin{document}

\title{High-temperature series for the
 bond-diluted Ising model in 3, 4 and 5 dimensions}
\date{\today}

\author{Meik Hellmund$^{1,3,}$}
\email{Meik.Hellmund@math.uni-leipzig.de}
\author{Wolfhard Janke$^{2,3,}$}
\email{Wolfhard.Janke@itp.uni-leipzig.de}
\affiliation{$^1$ Mathematisches Institut, Universit{\"a}t Leipzig,
Augustusplatz 10/11, D-04109 Leipzig, Germany\\
$^2$ Institut f{\"u}r Theoretische Physik, Universit{\"a}t Leipzig,
Augustusplatz 10/11, D-04109 Leipzig, Germany\\
$^3$ Centre for Theoretical Sciences (NTZ) of the Centre for
Advanced Study (ZHS), Universit{\"a}t Leipzig,\\
$^{~}\,\,\!$ Emil-Fuchs-Str.\ 1, D-04105 Leipzig, Germany}

\begin{abstract}
In order to study the influence of quenched disorder on second-order 
phase transitions,
high-temperature series expansions of the \sus\ and the free energy are 
obtained 
for the quenched bond-diluted Ising model in $d = 3$--5
dimensions. They are analysed using different extrapolation methods 
tailored to the expected singularity behaviours.
In $d = 4$ and 5 dimensions we confirm that the critical behaviour is 
governed by the pure fixed point up to dilutions near the geometric 
bond percolation threshold.
The existence and form of logarithmic corrections for the 
pure Ising model in $d = 4$ is
confirmed and our results for the critical behaviour of the diluted system
are in agreement with the type of singularity
predicted by renormalization group considerations. 
In three dimensions we find large crossover effects between the pure Ising,
percolation and random fixed point. We
estimate the critical exponent of the \sus\ to be  $\gamma =1.305(5)$ 
at the random fixed point. 

\end{abstract}
\pacs{\\
05.50.+q Lattice theory and statistics (Ising, Potts, etc.) \\
64.60.Fr Equilibrium properties near critical points, critical
          exponents\\
75.10.Hk Classical spin models\\
75.10.Nr  Spin-glass and other random models
}

\maketitle

\section{Introduction}

Since many years random Ising  models have served as paradigmatic 
systems in which the influence of quenched disorder may be studied through 
different techniques. 
In the present work we report results obtained by 
high-temperature series expansions.
Systematic series expansions \cite{domb3} for statistical physics models 
defined on a lattice 
provide an useful complement to field-theoretical 
renormalization group studies and large-scale numerical Monte Carlo
simulations. This is in particular true 
when studying phase transitions and critical phenomena 
of quenched, disordered systems. 

Series expansions techniques 
treat the quenched disorder average exactly and the infinite-volume limit 
is implicitly implied. Therefore one can obtain  exact 
results up to a certain order in the inverse temperature for many quantities.
Moreover, one can keep the disorder strength $p$ as well as the 
dimension $d$ as symbolic parameters and therefore analyse large regions of 
the parameter space of disordered systems. 
The critical part of the series
expansion approach lies in the extrapolation techniques which are used
in order to obtain information on the phase transition behaviour from
the finite number of known coefficients.
While for pure systems this usually works quite well,
one can question the use of these extrapolation techniques in disordered
systems, where the singularity structure of the free energy or susceptibility
may be very complicated, involving Griffiths-type singularities or logarithmic
corrections \cite{card99}. Yet our work indicates that at least for the
model we consider here, the extrapolation techniques are of comparable 
quality as for the case of the pure (no disorder) Ising model.

We consider in this paper 
the  Ising model on a hypercubic $d$-dimensional lattice ${\mathbb{Z}}^d$ 
with bond dilution as a realization of quenched and  uncorrelated
disorder. The pure model has a second-order phase transition for $d\geq2$,
and the upper critical dimension, where mean-field behaviour sets in, is
$d_u=4$. The influence of quenched disorder can be estimated by the
Harris criterion \cite{harris}: For $d>d_u=4$, the disorder is expected 
to change 
only non-universal quantities such as the transition temperature $T_c$.
At the upper critical dimension $d=4$, logarithmic or even more subtle
correction terms should appear, and in $d=3$, the phase transition of the 
disordered system should be governed by a new ``random'' fixed point since
for the pure model the critical exponent $\alpha$ of the specific heat is 
positive and disorder should hence be a relevant perturbation. 
Our high-temperature series analyses presented in this paper affirm 
this picture.

The rest of the paper is organized as follows: In Sect.~\ref{model} we first
briefly recall the model and some of its properties. Section~\ref{expansion}
is devoted to a description of the methods used by us for generating the 
series expansions, and Sect.~\ref{pade} first starts out with some remarks 
on the analysis techniques used. The main results are presented in the 
followings subsections, where we discuss our results for the random-bond 
Ising model in five (Sect.~\ref{5d}), four (Sect.~\ref{4d}) and three 
(Sect.~\ref{3d}) dimensions. Finally, Sect.~\ref{conclusions} contains 
our conclusions.

\section{The model}
\label{model}

The ferromagnetic disordered Ising model on hypercubic lattices 
$\mathbb{Z}^d$ is defined by the partition function
\begin{equation}
  \label{eq:1}
  Z \left(\{J_{ij}\}\right)  = \sum_{\{s_i\}} \exp \left(\beta \sum_{\langle ij\rangle }  
      J_{ij} s_i   s_j\right),
\end{equation}
where  $\beta = 1/k_B T$ is the inverse temperature, $J_{ij}$ are quenched
(non-negative) nearest-neighbour coupling constants, and the spins $s_i$ can
take on the two different values $\pm1$ . 
In our series expansion the combination  $v_{ij} = \tanh(\beta J_{ij})$ 
will be the relevant expansion parameter. 
Quenched disorder averages  $[\ldots]_{P(J)}$ such as the free energy  
\begin{equation}
  \label{eq:fe}
  -\beta F = [\ln Z]_{P(J)} 
\end{equation}
are taken over an uncorrelated bimodal distribution of the form
\begin{equation}
  \label{eq:bi}
  P(J_{ij}) = (1-p) \delta(J_{ij}-J_0) + p \delta(J_{ij}),
\end{equation}
which corresponds to bond dilution: With probability $p$, bonds are
effectively absent from the lattice, so that $p=0$ represents
the pure system. 
The expansion parameter for averaged quantities is usually taken as
$v = \tanh(\beta J_0)$.

\subsection{Thermal phase transitions}
\begin{figure}[tb]
  \centering
\includegraphics[scale=0.4,angle=-90]{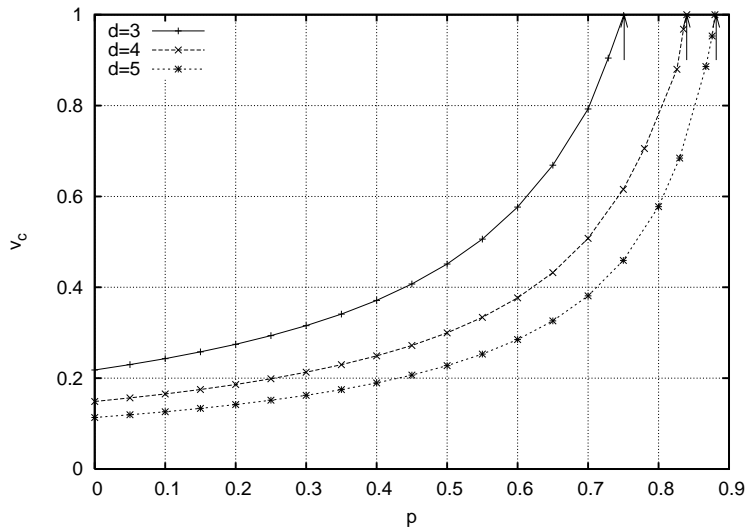}  
  \caption{Critical coupling $v_c$ as function of the dilution $p$ 
  for $d=3,4$, and 5 dimensions. 
The arrows on the top axis indicate the percolation thresholds.
}
  \label{fig:vc}
\end{figure}

The pure ($p=0$) model has for $d\geq2$
 a second-order phase transition from the disordered high-temperature phase
 to a ferromagnetic low-temperature phase. 
Bond dilution decreases $T_c$. Approaching 
the geometric bond percolation threshold $p=p_c$ the
critical temperature decreases to zero ($v_c\to1$) since above this value only
finite clusters of bonds are present and therefore no global 
ferromagnetic order is possible. 
Figure~\ref{fig:vc} gives an overview of the phase diagrams as calculated 
in this work from high-temperature series of the \sus. 
The influence of quenched disorder on the thermal fixed point is 
different above and below the upper critical dimension $d_u = 4$. 
For $d \geq 4$ 
the pure fixed point is, according to the Harris criterion \cite{harris}, 
 stable against the influence of disorder and one expects that the 
 renormalization group (RG) flow
goes along the critical line from the unstable percolation fixed point 
straight to the stable pure fixed point. The critical behaviour along this
line of phase transitions is governed by the pure fixed point.    
For $d=3$ the pure fixed point is unstable
and one expects a new random fixed point governing the critical behaviour
in the region $0<p<p_c$, attracting RG flow from both sides (pure and
percolation fixed point).
More specifically,  
 for $d=4$  random disorder changes the form of the 
logarithmic corrections to the Gaussian fixed point \cite{aharony76}, whereas
for $d=5$ Gaussian behaviour is expected for $p=0$ and $p>0$ as well. 

\subsection{Crossover to the percolation point}
\label{sec:perc}
The crossover behaviour of the diluted system near the percolation threshold
of the underlying lattice is well-understood 
\cite{coniglio81,aizenman87,staufferbook} and
independent of the dimension of the lattice.
The percolation critical point $p=p_c,\, T_c=0\,(v_c=1)$ is always unstable 
against thermal
fluctuations \cite{coniglio81}
 and plays for our system the role of the
 strong disorder fixed point \cite{igloi05}.

 The  crossover exponent 
 $\phi$ is exactly 1 and critical exponents approach their percolation values 
with a scaling behaviour of the form 
\begin{eqnarray}
  \chi &\sim& (p_c-p)^{-\gamma_p} f_{\chi}\left(\frac{e^{-2J_0/T}} 
{p_c-p}\right) \label{th1}\\
     &\sim& (p_c-p)^{-\gamma_p} f_{\chi}\left(\frac{1}{p_c-p} \times \frac{1-v}{1+v}\right),
     \label{th2}
\end{eqnarray}
where $f_{\chi}$ is a crossover-scaling function,
and the critical line near $v=1,\, p=p_c$ 
has the form 
\begin{equation}
  \label{th3}
 \frac{1-v}{1+v} \propto  (p_c-p).  
\end{equation}

\section{Star-graph expansion}
\label{expansion}

The star-graph expansion method has been reviewed many times in the
literature \cite{singh87} as well as  in our previous papers \cite{Hellmund:2003,
hellmund05a}. We will only  present some algorithmic details of our approach
and make some remarks comparing different techniques 
for series generation.

``Star graph'' is an old-fashioned  name for 1-vertex-irreducible graphs. 
For observables $F$ which allow a star-graph expansion 
(like the free energy and various inverse susceptibilities) 
this technique delivers a 
representation of $F$ on an 
infinite (e.g.\ hypercubic) lattice, 
\begin{equation}
  \label{eq:sg1}
  F(\mathbb{Z}^d) = \sum_G E(G;\mathbb{Z}^d)\, W_F(G),
\end{equation}
as sum over star graphs (1-vertex irreducible graphs) only. 
Here, $E(G;\mathbb{Z}^d)$ denotes the weak
embedding number of the graph $G$ in the given
lattice structure~\cite{martin74} and $W_F(G)$ can be recursively calculated
from the knowledge of $F(G)$ and $F(G')$ for all star subgraphs $G'\subset G$.

A property of eq.~(\ref{eq:sg1}) important for us is that the quenched disorder
average can be calculated on the level of single graphs as long as the
disorder is  uncorrelated. This is related to the fact that the method
considers and counts weak embeddings of the graph into the lattice, 
i.e., embeddings where different bonds of a graph are always mapped to 
different
bonds of the lattice. This is in contrast to the free embeddings used in
linked-cluster expansions. Here,
different vertices and bonds of the graph may correspond to the same vertex or
bond of the lattice.  
Therefore  in order to calculate the disorder average one
has to classify the free embeddings of a graph according to their
correspondence to 
embeddings of a collapsed graph (where some edges of $G$ are identified).
This is essentially the algorithm proposed in \cite{reisz99} where the
collapsed graphs are called "multiple-line graphs". 
This method has not yet been used for 
actual calculations of disordered systems,
 presumably due to its combinatorial complexity. 

For other series generation methods as, e.g., the
finite lattice method which holds the world record for the pure 3D Ising model
\cite{Arisue:2003} or methods using Schwinger-Dyson equations
\cite{Butera:99}, 
no practical generalization to disordered systems is known.

For the generation of graphs 
we employed the {\texttt nauty}
package by McKay \cite{mckay81} which makes very fast isomorphism tests by
calculating a canonical representation of the automorphism group of the graphs.
By this means, we
classified  all star graphs up to order 21 that can be
embedded in hypercubic lattices, see Table~\ref{tab:1}. 
For each of these graphs we
calculated their weak embedding numbers for $d$-dimensional hypercubic
lattices (up to order 17 for arbitrary $d$,  order  19 for
dimensions $d \leq 5$ and order 21 for $d \leq 3$). 
For this embedding count we implemented
a refined version of the backtracing algorithm by
Martin \cite{martin74}, making use of a couple of simplifications for 
bipartite hypercubic lattices ${\mathbb Z}^d$. After
extensive tests to find the optimal algorithm
for the innermost loop, the test for collisions in the embedding, we ended
up using optimized hash tables.

\begin{table}[tb]
\begin{center}
\caption{\label{tab:1}Number of star graphs with $N \geq 8$ links and
non-vanishing embedding numbers on ${\mathbb Z}^d$. For $N=1,4,6$, and 7
only a single star graph exists.}
  \begin{tabular}{l|*{14}{r}}
   \hline\hline
    order $N$&8&9&10&11&12&13&14&15&16&17&18&19&20&21\\
    \hline
    $\#$ in $d=3$  &2&3&8&9&29&51&136&306&856&2237&6431&18487&55302& 165730  \\
    $\#$ in $d=4$ and 5&2&3&8&9&29&51&142&330&951&2561&7622& 22688\\
   \hline\hline
  \end{tabular}
\end{center}
\end{table}

In order to calculate the contribution $W(G)$ of a graph $G$ of order $N$ 
to the 
\sus\ series, one needs the partition sum $Z(\{J_{ij}\}|G)$ 
and the matrix of all spin-spin
correlation functions $M_{kl}(\{J_{ij}\}|G) = \text{Tr } \delta_{s_k,s_l} e^{-\beta
  H({J_{ij})}}$ as polynomials in the $N$ variables $J_{ij}$. This is achieved by
using the cluster representation 

\begin{eqnarray}
  \label{eq:clu}
Z &\propto&  \sum_{C} 2^{c(C)} \prod_{\langle ij\rangle \in C} \frac{2 v_{ij}}{1 - v_{ij}},  \\
  \label{eq:clu2}
M_{kl} &\propto&  \sum_{C_{kl}} 2^{c(C)} \prod_{\langle ij\rangle \in C} \frac{2  v_{ij}}{1 - v_{ij}} ,
\end{eqnarray}
where the sum in eq.~(\ref{eq:clu}) goes over all clusters $C \subseteq G$ 
which have only vertices with an even number of bonds and 
in  eq.~(\ref{eq:clu2})
the sum is restricted to all clusters $C_{kl}\subseteq G$ in
which the vertices $k$ and $l$ have an odd number of bonds and all other have
an even number of bonds. The exponent $c(C)$ counts the number of connected
components of $C$. A graph with $N$ bonds gives rise to $2^N$ clusters:
every bond may be present or absent. The clusters can therefore be
 numerated by $N$-bit integers. These $2^N$ integers can be sorted in such a
 way that two consecutive numbers differ by exactly one bit 
(corresponding to the addition or deletion of one bond), an algorithm 
known as ``Gray code''~\cite{numrec}. 
This gives a simple algorithm for the calculation of the cluster sums
in eqs.~(\ref{eq:clu}) and (\ref{eq:clu2}):

\begin{itemize}
\item[--] Start with the full graph $C=G$,  store the coordination number 
  modulo 2 (``coordination bit'') for every vertex and 
      count the total number $r$ of odd vertices. 
Execute steps 2 and 3       from below for this cluster configuration. 
\item[--] Iterate over all other $2^N-1$ Gray codes $C$:
  \begin{enumerate}
    \item 
Calculate the next Gray code (cluster configuration). 
Compared to the previous code, exactly one bond was added or deleted. 
Invert the coordination bit of 
       the two involved vertices and calculate the change in $r$.
   \item If $r = 0$ then add $\prod_{n \in C} w_n$  to $Z$. The product has a 
         factor $w_n = 2 v_{ij}/(1 - v_{ij})$ for every non-zero bit $n$ in $C$.
    \item If $r=2$ then add   $\prod_{n \in C} w_n$ to $M_{kl}$
      where $k,l$ are the two odd vertices.
  \end{enumerate}
The prefactor $2^{c(C)}$ is taken into account
by monitoring the change in the number of connected components~$c(C)$ 
in each iteration step.
\end{itemize}

The further steps for the calculation of the \sus\ series are:
\begin{itemize}
\item Inversion of the $Z$ polynomial as a series in the $\{v_{ij}\}$
up to the desired order.
\item Averaging over quenched disorder,\\
$N_{kl}(G) = \left[ M_{kl}/Z \right]_{P(J)},$\\
  resulting in a matrix of polynomials in $p$ and $v$.
\item Inversion of the matrix $N_{kl}$  and subgraph subtraction,\\
 $W_\chi (G) = \sum_{k,l} (N^{-1})_{kl} - \sum_{G' \subset G} W_\chi(G')$.
\item Collecting the results from all graphs,\\
$1/\chi = \sum_G E(G;{\mathbb Z}^d)\; W_\chi(G)$.
\end{itemize}
All calculations are done in arbitrary-precision integer arithmetic using the
open source library GMP. For the polynomial arithmetic we developed our own
optimized C\raise2pt\hbox{\tiny++}  template library using a degree-sparse 
representation of polynomials as linked lists~\cite{Knuth}.
The calculations took around two CPU years on an Opteron Linux Cluster.

\section{Series analyses}
\label{pade}

The estimate of critical parameters from a high-temperature series 
involves extrapolation from a finite number of exactly known coefficients 
to the asymptotic form of the function. Many such extrapolation techniques
have been developed and tested for different series and are comprehensively 
reviewed in \cite{guttmann89}. 
These extrapolation techniques are not rigorous. They make some assumptions
about the expected form of the singularity at the critical temperature.    
Field theoretic techniques like the $\epsilon$-expansion have similar problems.

Usually, error estimates rely on the scatter of the results of extrapolations
with different parameters (like $[N/M]$ Pad\'e approximants  for different 
values of $N$ and $M$). This may seriously underestimate systematic errors
coming from wrong assumptions about the structure of the singularity. 

In order to get a reliable picture, we will take into account several
criteria, such as 
\begin{itemize}
\item convergence of the analysis,
\item scatter of different approximants, 
\item number of defective approximants,
\item agreement between different extrapolation methods.  
\end{itemize}

The basic methods we use are DLog-Pad\'e approximation and inhomogeneous
differential approximants (IDA) \cite{guttmann89}. 
In order to analyze confluent nonanalytic
and logarithmic corrections,  these methods are applied to suitably
transformed forms of the series. The parameters of these transformations are 
fine-tuned according to the criteria listed above, a technique 
pioneered in \cite{adler91}.

We obtain from the star-graph technique series in two variables $v$ and $p$. 
Since the algorithm involves inversion of polynomials (and of matrices of
polynomials) one has to find a consistent truncation criterion for the
resulting series. The crucial observations are that a) the coefficient matrix 
$a_{mn}$ of the 
resulting series  
is triangular (cf.~App. \ref{app}), i.e., the series is of the form 
\begin{equation}
  \chi(p,v) = \sum_{n=0}^\infty  \,   \sum_{m=0}^n a_{mn} q^m v^n\qquad  \text{where}\qquad q= 1-p, 
\end{equation}
and b) the contribution of a graph of order $N$ starts with terms of order 
$q^N v^N$. 
This allows different consistent truncation schemes. In scheme~A 
we calculate series which are correct up to a given order $N$ in $v$,
i.e., we calculate all non-vanishing coefficients $a_{mn}$ with  $n\leq N$. 
In scheme~B, we calculate all non-vanishing coefficients $a_{mn}$ which
satisfy $m+n\leq 2N$. 

Both schemes are useful in different regions of the
parameter space. 
At small and medium disorder $p$ we approach the 
critical point by extrapolating in $v$ for constant values of $p$. 
Therefore we use the series calculated according to truncation scheme~A 
in the form 
\begin{equation}
  \label{eq:x1}
\chi(v|p={\rm const.}) \propto   \sum_{n=0}^N \, \left(  \sum_{m=0}^n a_{mn} q^m \right) v^n
= \sum_{n=0}^N c_n v^n.  
\end{equation}
Near the percolation threshold $p\lesssim p_c$ this gives less and less 
satisfactory results. As eq.~(\ref{th3}) and Fig.~\ref{fig:vc} show, 
$p={\rm const.}$-lines 
 are reaching the critical line at a smaller and smaller angle. In this
 region, it is better to extrapolate along the locus $w=q/v= {\rm const.}$
This is achieved by taking the series coefficients calculated with scheme~B:
\begin{equation}
  \label{eq:x2}
\chi(v|w={\rm const.}) \propto   \sum_{m+n\leq2N}  a_{mn} w^m v^{n+m} 
= \sum_{n=0}^{2N} d_n v^n.  
\end{equation}

The analysis of scheme~B series brings much better results than scheme~A 
in the highly diluted region $p\lesssim p_c$. But  the quality of the analysis is 
still lower (e.g., the scatter of different Pad\'e approximants is larger) than in
the regions with small or medium dilution.  This is not surprising since 
we can not construct from our data a series consistently 
truncated in $p$. This would be necessary in order to  extrapolate along  the  
locus $v={\rm const.}$ and study the percolation transition directly. 
Nonetheless, a reliable estimate of the transition temperature $v_c$ is 
possible up to $p\approx p_c$, giving the phase diagram in Fig.~\ref{fig:vc}.

\subsection{Five dimensions}
\label{5d}

The analysis of the 19th order  series for the
\sus\  of the 5D bond-diluted Ising
model is a nice warmup in order to get an impression of the quality of
different analysis methods. Unbiased DLog-Pad\'e and inhomogeneous differential
approximants (IDA) along different $p={\rm const.}$ loci 
 give  values of $\gamma$ between $1.02$ and $1.04$ for a wide range
of dilutions $p=0\ldots0.7$ with a slightly increasing tendency.

This effect -- an overestimate of $\gamma$ in a na\"\i ve analysis not taking into 
account confluent corrections -- is well-known
from the 3D pure Ising model. A good illustration is Fig.~2 in \cite{campo02}:
whereas the results for many different theories in the Ising universality
class ($\phi^4$ etc.) converge to $\gamma\approx1.237$, even 25th order series for the
Ising model give $\gamma\approx 1.244$. The remedy is the use of analysis methods
tailored for confluent nonanalytic corrections,

\begin{equation}
  \label{eq:conf}
  f(v) \sim A (v_c-v)^{-\gamma} \left[1 + A_1 (v_c-v)^{\Delta_1} + \ldots\right], 
\end{equation}
 such as the methods from \cite{adler91} called 
M1 and M2. The method M1 uses DLog-Pad\'e approximants to 
\begin{equation}
  \label{eq:m1}
  F(v) = (v_c-v) \frac{df}{dv} -\gamma f(v),
\end{equation}
    which has a pole at $v_c$ with residue $\Delta_1 -\gamma$. For a given trial value
    of $v_c$ the graphs of $\Delta_1$ versus input $\gamma$ are plotted for different
    Pad\'e approximants and by adjusting $v_c$ a point of optimal convergence is
    searched.  

The M2 method starts with a transformation of the series in $v$ into a series
in $y=1-(1-\frac{v}{v_c})^{\Delta_1}$, and then Pad\'e approximants to
\begin{equation}
  \label{eq:M2}
  G(y) = \Delta_1 (1-y) \frac{d}{dy} \ln F(y)
\end{equation}
are calculated which should converge to $\gamma$ as $y\to1$, i.e. $v \to v_c$.
These methods are especially useful when taken as biased
    approximants with a given value of $\gamma$ or $\Delta_1$ as input. 

\begin{figure}[t]
  \centering
  \includegraphics[scale=0.4,angle=-90]{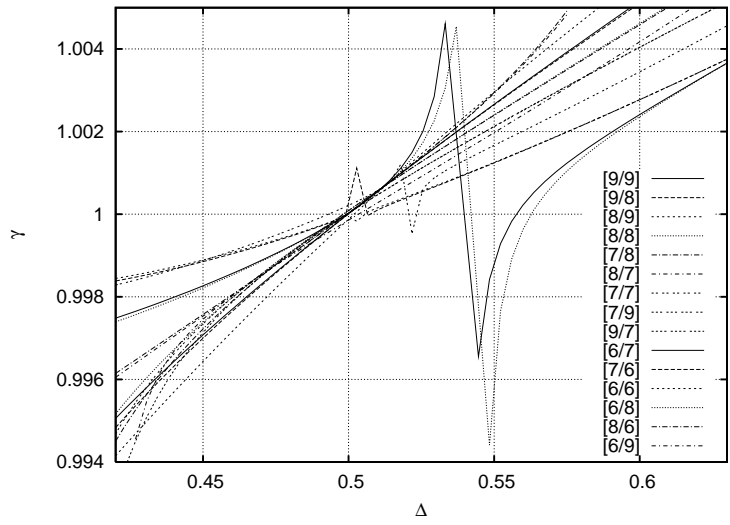}
  \caption{M2 analysis ($\gamma$ as function of $\Delta_1$) for the 
  $p=0.5$ diluted 5D Ising model at $v_c = 0.227\,498$ based on 15 different
  $[N/M]$ Pad\'e approximants.}
  \label{fig:is5d}
\end{figure}

Using the M2 method  
we find over a large range of the disorder strength 
$p=0\ldots0.7$ excellent convergence of the Pad\'e approximants with a nearly
$p$-independent value of $\Delta_1$. 
 As an example,
Fig.~\ref{fig:is5d} shows a M2 plot at dilution $p=0.5$ from which we read off
$\gamma = 1.001(1), \Delta_1=0.51(2)$ and determine $v_c = 0.227\,498$. 
Our overall estimate for $\Delta_1$ for all  
disorder strengths  $p=0\ldots0.7$ is $\Delta_1=0.50(5)$. 
The results of the M1 method are compatible with this but
have poorer convergence.  

For the pure ($p=0$) 
5D Ising model our critical values obtained from the 19th order series
with the M2 analysis biased with $\gamma=1$ are
$v_c = 0.113\,425(3), \Delta_1 = 0.50(2)$  
As usual, the errors are estimates of the scatter of different Pad\'e
approximants. This should be compared with the value from Monte Carlo
(MC) simulations \cite{binder99} of $\beta_c = 0.113\,915\,2(4)$, 
i.e. $v_c= 0.113\,425\,0(4)$, which is in perfect agreement.   

\subsection{Four dimensions}
\label{4d}

Four is the upper critical dimension of the Ising model. 
Without impurities the scaling behaviour of the \sus\ and specific heat 
is thus expected to exhibit at the critical point
logarithmic corrections. With $t \equiv |1 - T/T_c|$, the precise form
reads as \cite{kenna1,kenna2,kenna3,balle4d}
\begin{eqnarray}
  \label{eq:sus4dp}
  \chi &\sim& t^{-1} |\ln t|^{1/3},\\
  C  &\sim&  |\ln t|^{1/3}, \label{eq:c4}
\end{eqnarray}
where a non-singular background term in $C$ has been omitted.
For the disordered case it was shown long ago \cite{aharony76} 
by a RG analysis 
that the critical behaviour is modified to take the form 
\begin{eqnarray}
  \label{eq:sus4aha}
  \chi &\sim& t^{-1} \exp\left[D|\ln t|^{1/2}\right],\\
 C &\sim&   -|\ln t|^{1/2} \exp\left[-2D|\ln t|^{1/2}\right], \label{eq:c41}
\end{eqnarray}
where $D=\sqrt{6/53} \approx 0.336$. The crossover to mean-field
behaviour as well as higher loop corrections \cite{geldart93}
modify this unusual exponential term by  factors of the form $|\ln t|^{\Delta_1}$.
Notice that the expression in eq.~(\ref{eq:c41}) approaches zero for $t\to0$ 
such that $C$ is no longer logarithmic divergent but
stays finite in the presence of disorder.

\subsubsection{Pure Ising model}

Due to the logarithmic corrections in eqs.~(\ref{eq:sus4dp}) and (\ref{eq:c4})
a special treatment of the series is needed. The analysis of logarithmic 
corrections of the general form
\begin{equation}
  \label{eq:log}
  f(v) \sim (v_c-v)^{-\gamma} |\ln (v_c-v)|^{\delta}
\end{equation}
is possible by \cite{adler81} calculating approximants for 
\begin{equation}
  \label{eq:log2}
F(v) = (v_c-v) \ln(v_c-v) \left[\frac{f'(v)}{f(v)}-\frac{\gamma}{v_c-v}\right],
\end{equation}
where one expects for singularities of the form (\ref{eq:log}) that 
$\lim_{v\to v_c} F(v) = \delta$. In what follows we call this the M3 method.

\begin{figure}[t]
  \centering
  \includegraphics[scale=.4,angle=-90]{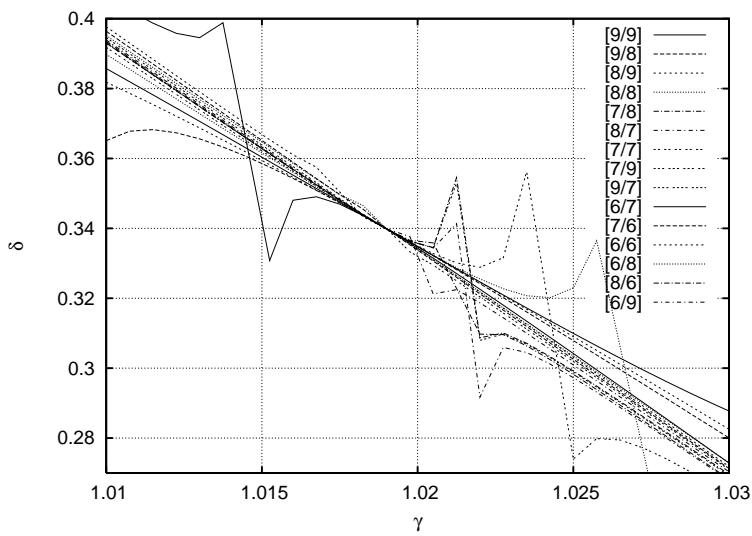}
  
  \caption{Logarithmic exponent $\delta$ as function of the leading exponent
$\gamma$ for the pure 4D Ising model via unbiased 
M3 approximants at $v_c=0.148\,607$. A pronounced clustering of the different
$[N/M]$ Pad\'e approximants around the point $\gamma \approx 1.019$,
$\delta \approx 0.34$ is clearly observed.}
  \label{fig:4dp2}
\end{figure}

\begin{figure}[t]
  \centering
  \includegraphics[scale=.4,angle=-90]{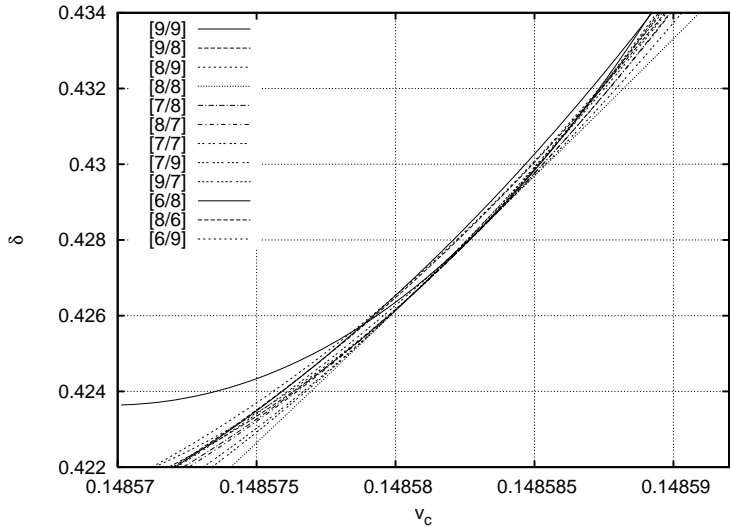}
  \caption{Logarithmic exponent $\delta$ as function of $v_c$ for the pure 4D Ising model 
via M3 approximants biased to $\gamma=1$. From the clustering of the $[N/M]$ Pad\'e 
approximants one reads off $v_c \approx 0.148\,583$, $\delta \approx 0.429$. When
assessing the clustering property notice that the $y$-scale is 10 times finer than
in Fig.~\ref{fig:4dp2}.}
  \label{fig:4dp1}
\end{figure}

The estimates for the pure 4D Ising model based on our 19th order series are: 
a) unbiased estimate, Fig.~\ref{fig:4dp2}:
$\gamma=1.019(1), \delta=0.34(1), v_c=0.148\,607(10)$ 
and b) biased to $\gamma=1$, Fig.~\ref{fig:4dp1}:
$\delta=0.429, v_c=0.148\,583(3)$.
This should be compared to the analysis in \cite{staufadl}
  of a 17th order series \cite{gaunt79} which gave $v_c=0.148\,588$ 
and to MC simulations \cite{bittner02} which found
$\beta_c =  0.149\,697(2),$ i.e. $v_c = 0.148\,589(2)$.
While the agreement of the estimates for $v_c$ is excellent, this
analysis also shows how difficult it is to obtain reliable
estimates for the critical exponent $\delta$ of the logarithmic correction
term.

\subsubsection{Bond dilution}

An extensive MC study of the {\em site\/}-diluted 4D model 
was done in \cite{balle4d}. The authors found data compatible with the
theoretical prediction of a Gaussian fixed point with logarithmic corrections,
but a precise fit of the logarithmic corrections was not possible. 

\begin{figure}[tb]
  \centering
  \includegraphics[scale=0.4,angle=-90]{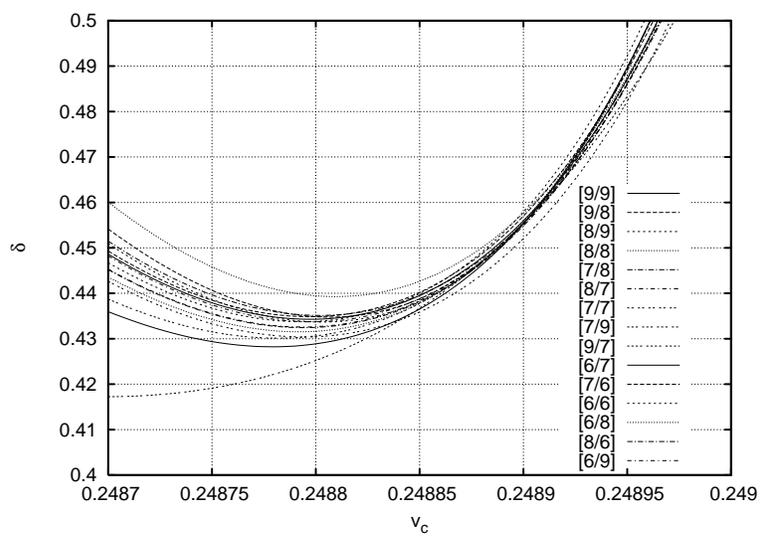}
  \caption{Logarithmic exponent $\delta$ as function of $v_c$ for the $p=0.4$ diluted 4D model, 
results of a M3 analysis.}
  \label{fig:4dm3}
\end{figure}

\begin{figure}[tb]
  \centering
  \includegraphics[scale=0.4,angle=-90]{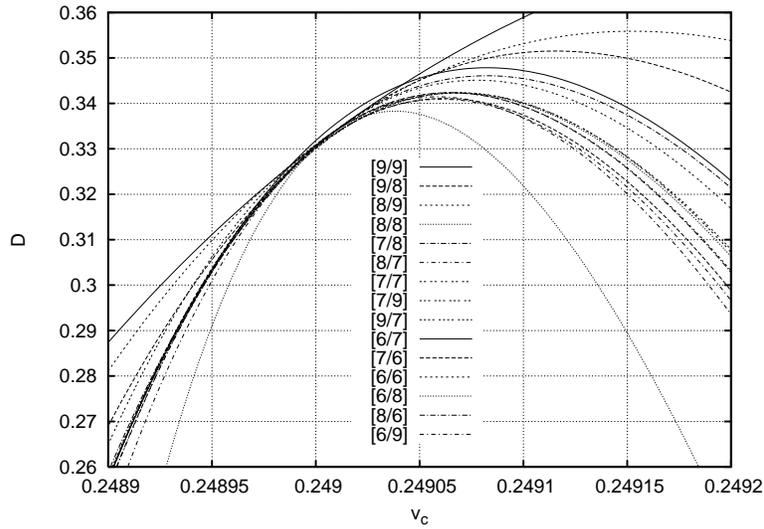}
  \caption{Parameter $D$ as function of $v_c$ for the $p=0.4$ diluted 4D model, 
results of a M3a analysis.}
  \label{fig:4dm3a}
\end{figure}

A modification of the M3 analysis method which we refer to as M3a is able 
to take into account the specific form of eq.~(\ref{eq:sus4aha}): For 
\begin{equation}
  \label{eq:sus41}
  f(v) \sim (v_c-v)^{-\gamma} \exp \left[D|\ln(v_c-v)|^{1/2}\right] 
\end{equation}
we calculate approximants to 
\begin{equation}
  \label{eq:sus42}
  F(v) \sim (v_c-v) (-\ln(v_c-v))^{1/2} \left[\frac{f'(v)}{f(v)}-\frac{\gamma}{v_c-v}\right]
\end{equation}
which should behave as $\lim_{v\to v_c} F(v) = D/2$.
We made an extensive analysis of our 19th order \sus\ series assuming the 
three forms eqs.~(\ref{eq:conf}), (\ref{eq:log}) resp.\ (\ref{eq:sus4dp}) and (\ref{eq:sus41})
resp.\ (\ref{eq:sus4aha}).
It turns out that, while form~(\ref{eq:conf}) is not acceptable,
 both forms of logarithmic corrections allow a fit of the
series data, both with a disorder dependent exponent $\delta$ or $D$, respectively. 
The quality of the M3a fits is, however, better than for the M3 fits. 
This is demonstrated in  Figs.~\ref{fig:4dm3} and \ref{fig:4dm3a} where 
equal ranges of $v_c$ $(\Delta v_c = 0.0003)$ and of the exponent 
$(\Delta \delta = \Delta D = 0.1)$  are shown.  
 
\begin{figure}[htb]
  \centering
  \includegraphics[scale=0.4,angle=-90]{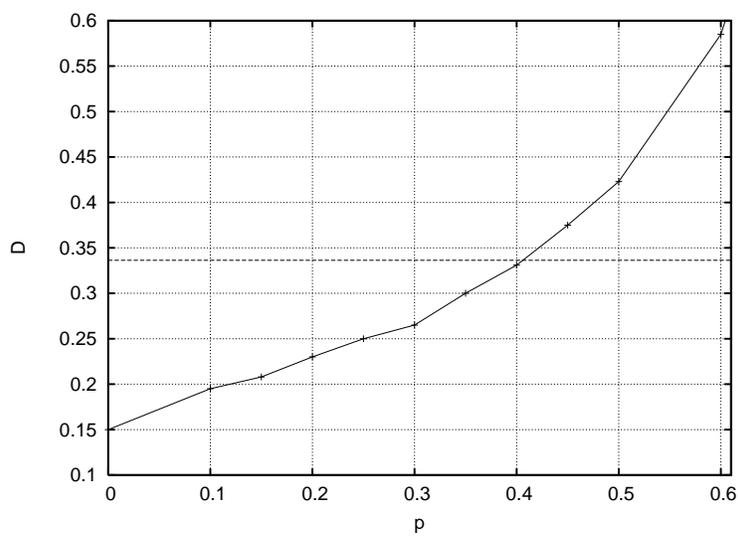}
  \caption{Parameter $D$ as function of $p$ for the  diluted 4D model, 
results of a M3a analysis. The horizontal line at $D=0.336\dots$ shows the 
predicted value $\sqrt{6/53}$.}
  \label{fig:4dm3b}
\end{figure}

We interpret this as an indication for the validity of the RG prediction
eq.~(\ref{eq:sus4aha}).
Figure~\ref{fig:4dm3b} shows the dilution dependency of our estimates for the
critical parameter $D$. It illustrates the difficulty in differentiating
between a behaviour of the form eq.~(\ref{eq:log}) and the form 
eq.~(\ref{eq:sus41}) 
by giving a non-zero result for $D$ at $p=0$.
It is nevertheless quite impressive that we can see at all  such weak correction 
terms to the leading singularity and even estimate the parameters.  

For the specific heat, one expects the weak logarithmic singularity in the pure
case [eq.~(\ref{eq:c4})] to be washed out by the disorder. 
An analysis of the free energy series $[F(v)]_{P(J)} = \Sigma a_i(p) v^i $ is 
usually 
more difficult and gives less satisfactory results compared to the \sus\
series. One looses two orders in $v$ by calculating  $C=\beta^2 \frac{d^2F}{d\beta^2}$  
and, more importantly, on bipartite lattices the series for $F(v)$  includes only 
even powers of
$v$. So it has to be considered as a much shorter series in $v^2$. In four 
dimensions we calculated $F(v,p)$ up to order 18, which gives a series for
$C(v^2)$ of order 8 in $v^2$.

Another difficulty is that the non-singular background terms are more
influential in this case. Therefore DLog-Pad\'e approximants show very poor
convergence and one has to use inhomogeneous differential approximants (IDA)
which are able to take background contributions into account.  By this method,
polynomials $P_N(v), Q_M(v)$ and $R_L(v)$ (of order $N, M$ and $L$, 
respectively) are determined such that 
\begin{equation}
  \label{eq:ida}
  P_N(v) \hat D f(v) + Q_M(v) f(v) + R_L(v) = o(v^{(N+M+L)} ),
\end{equation}
where many different triples $(N,M,L)$ with $N+M+L\leq\text{(order of~} f)$ 
and two different variants of the differential operator $\hat D$, 
either $\hat D_1=v d/dv$ or $\hat D_2=d/dv$, 
are used. The critical point $v_c$ is then given by the smallest positive real
root of $P_N(v)$ and the critical exponent by $Q_M(v_c)/P_N'(v_c)$.

\begin{figure}[tb]
  \centering
  \includegraphics[scale=0.4,angle=-90]{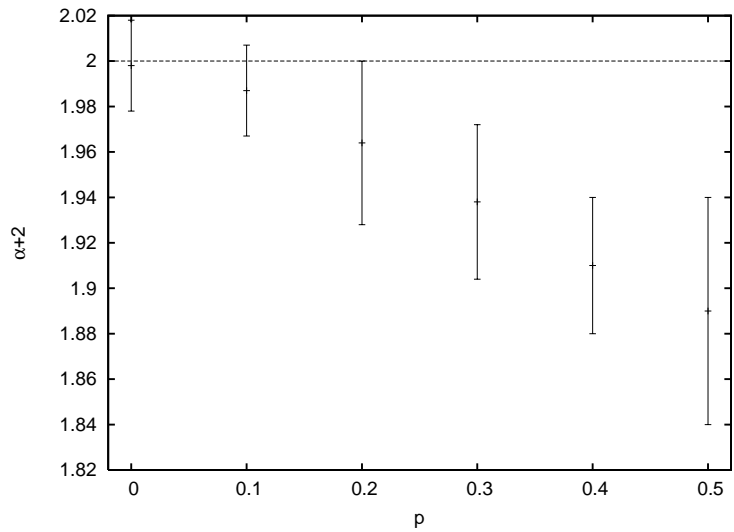}
  \caption{Critical exponent of $\partial^4f(v)/ \partial v^4$ 
 as function of $p$ for the  diluted 4D model.}
  \label{fig:4dC}
\end{figure}

We 
applied IDAs tailored to power-like singularities $\sim t^{-\alpha-2}$ to the fourth
derivative 
$\partial^4f(v)/ \partial v^4$. The results are shown in Fig.~\ref{fig:4dC}.
In the pure
case we find a result consistent with $\partial^4f(v)/ \partial v^4 \sim (v_c-v)^{-2}$
indicating a logarithmic singularity in $C\sim  \partial^2f(v)/ \partial v^2$. 
At nonzero values of the dilution parameter $p$ the singularity of the 
fourth derivative of $f$ is clearly weaker than $(v_c-v)^{-2}$, indicating 
 the absence of a divergence in the specific heat in the disordered case.

\subsection{Three dimensions}
\label{3d}

Let us finally turn to the physically most important case of three
dimensions. The quest for a determination of the properties of the expected
random fixed point in the 3D disordered Ising model is already rather long. 
A comprehensive compilation of results can be found in
\cite{folk99, folk01a} and \cite{peliset}, 
showing a wide scatter in the critical exponents
 of different groups, presumably due to large crossover effects.
Recent MC simulations \cite{balles98, holov05a,berche04, berche05} 
provide evidence for the  random fixed point but also show   large
  crossover effects due to the interference with the pure fixed
  point for $p \to 0$ and with the percolation fixed point for
  $p \to p_c$ (recall the discussion in Sect.~\ref{sec:perc}). 

Early attempts using series expansions \cite{rap1,rap2} already indicated that
the series are slower converging and more difficult to analyse than in the pure
case. A review of earlier work can be found in \cite{stinchcombe83}.
Series analysis in crossover situations is, in fact, extremely
difficult. If the parameter $p$ interpolates between regions 
governed by different fixed points, the
exponent obtained from a finite number of terms of a 
series expansion must cross somehow between
the two universal values, and does this usually quite slowly~\cite{Hellmund:2005b}.
The mere existence of a plateau in $\gamma(p)$,
however, is an indication that here a truly different critical behaviour is seen.

\begin{figure}[t]
  \centering
  \includegraphics[scale=0.4,angle=-90]{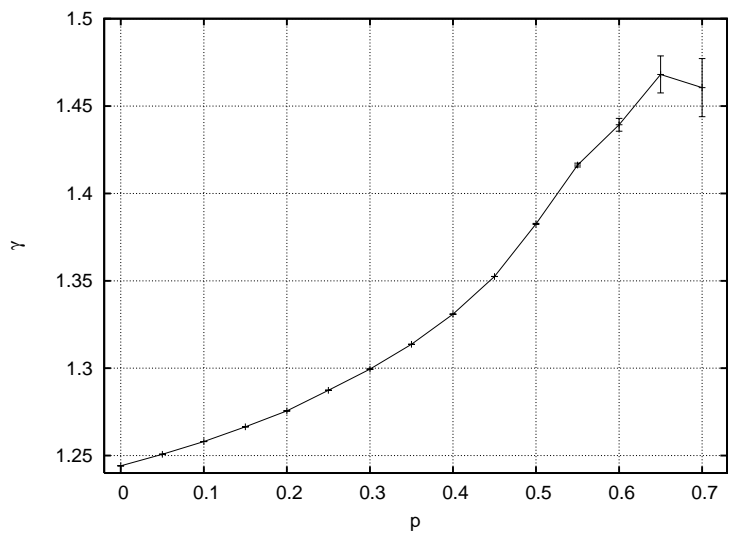}
  \caption{Critical exponent $\gamma$ as function of $p$ for the  
  diluted 3D model obtained from about 300 DLog-Pad\'e and inhomogeneous 
  differential approximants for each dilution value.
  }
  \label{fig:3gam}
\end{figure}

Our results are obtained from a large number of DLog-Pad\'e and inhomogeneous
differential approximants (around 300 for every dilution value) applied to the
21th order susceptibility series $\chi(p,v)$ compiled in App.~A. The resulting
estimates for the critical exponent $\gamma$ shown 
in 
Fig.~\ref{fig:3gam}, however, do not exhibit any sign for a 
plateau.

\begin{figure}[t]
  \centering
  \includegraphics[scale=0.4,angle=-90]{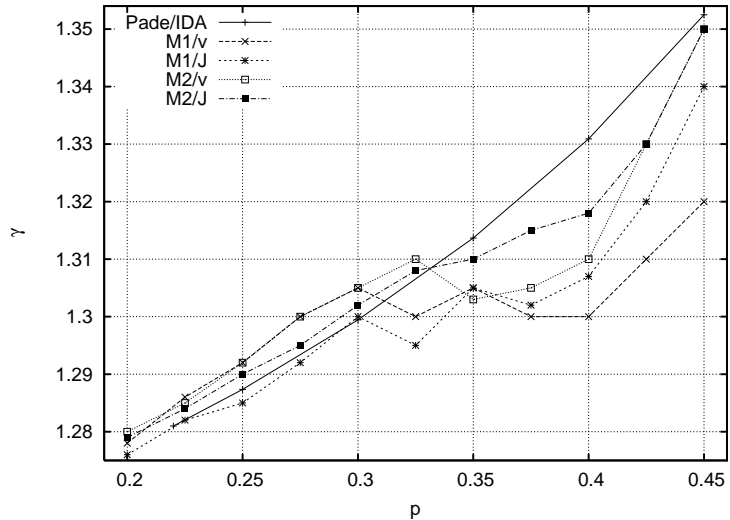}
  \caption{Critical exponent $\gamma$ as function of $p$ for the diluted 3D
    model obtained from series analyses with methods M1 and M2 (see text
    for details). The data of Fig.~\ref{fig:3gam} (Pade/IDA) are also shown 
    for comparison.
    }
  \label{fig:3gam1}
\end{figure}

Since confluent corrections are essential to understand crossover
situations, we also performed a careful analysis using the M1 and M2 methods which both
take such corrections explicitly into account, see Fig.~\ref{fig:3gam1}.
  This figure shows results using both $v=\tanh (\beta J_0)$ as well as the 
  coupling $\beta J_0$ as
expansion variables (denoted by, e.g., M1/$v$ and
M1/$J$). They give an indication of a
plateau around $p=0.3\dots0.4$, suggesting the presence of the random fixed
point with corresponding value of the critical
exponent $\gamma=1.305(5)$. 
Here, as usual in extrapolation techniques, the error is 
estimated from the scatter of different approximants and we are unable to
give an estimate of the systematic error of the extrapolation.
Our high-temperature series estimate is at least compatible
with MC results for site and bond
dilution 
 \cite{berche04,Calabrese1,balles98}
which cluster quite
sharply around $\gamma_{\rm MC} = 1.34(1)$. Field-theoretic RG
estimates 
 \cite{folk99,Pelissetto1} 
favor slightly smaller exponents
of $\gamma_{\rm RG} = 1.32$ -- 1.33,
while experiments 
 \cite{Birgeneau0,Birgeneau1,Belanger1}
report values between $\gamma_{\rm exp} = 1.31$ -- 1.44; for a more
detailed compilation, see e.g., Table~1 in Ref.~\cite{berche05}.
Since it would be extremely demanding to further extend the series expansions
in the disordered case,
better series analysis methods for the case of crossover situations would 
be clearly desirable. 

\section{Summary}
\label{conclusions}

We successfully applied the method of high-temperature series expansion
to the bond-diluted Ising model in several dimensions. The computational
effort is vast and increases faster than exponentially with the order 
of the expansion. But 
the extensive set of combinatorical data we generated on the way, 
such as the list of star graphs
and their embedding numbers into hypercubic lattices, has a large number of
potential applications, for example to models with other kinds of
uncorrelated disorder like  spin glasses. 
On the other hand, Monte Carlo simulations of systems with 
quenched disorder require an enormous
amount of computing time because many realizations have to be simulated for
the quenched average. For this reason it is hardly possible to scan a whole
parameter range. Using the method of star-graph expansion we obtain
 this average exactly. Since the  relevant parameters (degree
of disorder $p$, spatial dimension $d$,  etc.)
are kept as symbolic variables, we can easily study large regions of the
phase diagram.  

Table~\ref{tab:sum} summarizes the main results of this work.  
Our estimates of the critical temperature for the 
4- and 5-dimensional model are comparable in precision (as well as
consistent) with the available Monte Carlo data for the pure case.
For the bond-diluted model in 4 and 5 dimensions we demonstrated 
universality. The critical exponent $\gamma$ keeps its Gaussian 
value $\gamma=1$ up to the percolation threshold of the
underlying lattice, despite the fact that the upper critical dimension
of the percolation transition is six.
In 5 dimensions we also show that the exponent $\Delta_1$  of the confluent
correction is universal for a large range of the dilution parameter $p$.
In 4 dimensions we confirm the RG prediction that the weak logarithmic
divergence of the specific heat of the pure model disappears in the
disordered case  due to logarithmic corrections of a special
kind. We confirm the presence of these special corrections also in the case
of the \sus. We furthermore determine the exponent of the logarithmic correction
in the pure case with satisfying precision.

The case of 3 dimensions is, in a sense, the most difficult to analyze.
We clearly see that the pure fixed point is unstable and does not describe
the random system. Extrapolation methods not taking into account confluent
corrections give a value for $\gamma$ which changes monotonously with
$p$. By using methods tailored to the consideration of confluent corrections
we identify a plateau in the curve $\gamma(p)$  with  critical exponent
$\gamma=1.305(5)$ which we see as evidence for the random critical point.

\begin{table}[hbt]
  \centering
\setlength{\tabcolsep}{4mm}
  \begin{tabular}{|c|c|r|l|l|}
 \hline 
 \hline 
 \multicolumn{2}{|c|}{model}  & \multicolumn{1}{c|}{$v_c$} 
& \multicolumn{1}{c|}{$ \gamma $}  & confluent correction \\ \hline 
 & pure & 0.113\,425(3) &1.001(1) & power-like, $\Delta_1=0.50(2)$ \\ \cline{2-5}
\raisebox{1.5ex}[0pt]{5D}&disordered & & 1.001(1)& 
power-like, $\Delta_1=0.50(5)$\\ \hline
  & pure&  0.148\,607(3)& 1.019(20) & logarithmic, $\delta=0.34(2)$  \\ \cline{2-5}
 \raisebox{1.5ex}[0pt]{4D} & disordered && 
\multicolumn{2}{|c|}{form eq.~(\ref{eq:sus4aha}) confirmed 
} \\ \hline
3D & disordered && 1.305(5) &\\ 
 \hline  
 \hline  
  \end{tabular}
  \caption{Summary of results.  
In three dimensions, longer series for the pure Ising model exist,
therefore we do not quote here our results for $v_c$ or $\gamma$.   
}
  \label{tab:sum}
\end{table}

\begin{acknowledgments}
 We thank Joan Adler for introducing us into the art and science
of series extrapolation techniques. 

    Support by  DFG grant No.~JA 483/17-3 and partial support from the
  German-Israel-Foundation under
          grant No.~I-653-181.14/1999 is gratefully acknowledged.
\end{acknowledgments}

\appendix
\section{Coefficients of the 3D \sus}
\label{app}
The following table of coefficients gives the complete information 
for calculating the susceptibility of the 3D bond-diluted Ising model
up to the order $v^{21}$ for any dilution $p$.

\begin{turnpage}
\squeezetable
\begin{table}
\hspace*{-3.7cm}
\caption{Table of coefficients $a_{mn}$ 
of $\chi(p,v)=1 + 6 q v + 30 q^2 v^2 +150q^3 v^3 + \sum a_{mn} q^m v^n$ up to order $n=21$, where
  $q=1-p$.  In the
  table, $m$ is the column number starting from 4 and $n$ is the row number
  starting from 4, so 
$\chi=1 + 6 q v + 30 q^2 v^2 +150q^3 v^3 +726 q^4 v^4 -24 q^4 v^5 +
3534 q^5 v^5 \ldots $.}
\hspace*{-3.7cm}
 \begin{tabular}{rrrrrrrrrrrrrrrrrr}
\hline
\hline
 726 \\
 $-24$ & 3534 \\
 $-24$ & $-192$ & 16926 \\
 $-24$ & $-192$ & $-1608$ & 81318 \\
 0 & $-192$ & $-1608$ & $-10464$ & 387438 \\
 24 & 0 & $-1608$ & $-10536$ & $-67320$ & 1849126 \\
 24 & 192 & $-264$ & $-9744$ & $-67632$ & $-395328$ & 8779614 \\
 24 & 192 & 1080 & $-240$ & $-60912$ & $-397704$ & $-2299560$ &
41732406 \\
 0 & 192 & 1344 & 8400 & $-1440$ & $-339936$ & $-2295744$ &
$-12766944$ & 197659950  \\
$-24$ & 0 & 1608 & 10296 & 51480 & 26544 & $-1886928$ &
$-12680496$ & $-70404720$ & 936945798  \\
  $-24$ & $-192$ & 264 & 10032 & 64560 & 341568 & 259656 &
$-9915696$ & $-69162048$ & $-377522064$ & 4429708830 \\
  $-24$ & $-192$ & $-1080$ & $-1704$ & 60024 & 427920 & 2062368 &
2482464 & $-51644200$ & $-367148472$ & $-2014331904$ & 20955627110  \\
  0 & $-192$ & $-1080$ & $-12144$ & $-14448$ & 360192 & 2493600 &
12550416 & 17926128 & $-259622976$ & $-1931961792$ & $-10550435184$ & 98937385374 \\
  24 & 0 & $-1080$ & $-13440$ & $-80928$ & $-132024$ & 1840776 &
14790144 & 73051512 & 126567264 & $-1293631728$ & $-9980137536$ & $-55050628008$ &
467333743110  \\
  24 & 192 & 0 & $-8544$ & $-95040$ & $-569760$ & $-1214880$ &
9797904 & 82573800 & 420942768 & 807789264 & $-6273975792$ & $-51221501136$ &
$-283516855968$ & 2204001965006  \\
  24 & 192 & 1080 & 5832 & $-60888$ & $-705216$ & $-3910368$ &
$-8858136$ & 46862760 & 457439184 & 2361075624 & 5069434800 & $-30136593768$ &
$-259361429784$ & $-1455780298776$ & 10398318680694  \\
 0 & 192 & 1080 & 18912 & 36720 & $-437952$ & $-4512600$ &
$-25012512$ & $-63580104$ & 220823568 & 2449336680 & 13097561328 & 30177202248 &
$-141380350848$ & $-1306684851840$ & $-7403140259952$ & 48996301350750  \\
  $-24$ & 0 & 1080 & 24744 & 129816 & 283752 & $-2272584$ &
$-28419456$ & $-158605280$ & $-422656608$ & 936811968 & 12971851368 & 71258617752
& 177314558064 & $-655481735280$ & $-6514909866600$ & $-37556614417032$ &
230940534213046\\
\hline
\hline
\end{tabular}
\end{table}
\end{turnpage}

\newpage\

\end{document}